# Magnetically modulated fluorescent probes in turbid media

Zhiqiang Yang, KhanhVan T. Nguyen, Hongyu Chen and Jeffrey N. Anker[*]

Clemson University Department of Chemistry, 371 Hunter Laboratories, Clemson, SC. 29634

* Corresponding author: janker@clemson.edu

**Abstract**

Magnetically modulated optical nanoprobes (MagMOONs) were used to detect and distinguish probe fluorescence from autofluorescent backgrounds in turbid media. MagMOONs are micro/nano-sized particles with magnetically controlled orientation and orientation-dependent fluorescence. These probes blink when they rotate in response to rotating external magnetic fields. This blinking signal can be separated from backgrounds enabling spectrochemical sensing in media with strong autofluorescence. We explore the effect of scattering on MagMOON fluorescence. Turbid media reduce the modulated MagMOON signal due to a combination of attenuation of fluorescence signal and reduction in contrast between "On" and "Off" states. The blinking MagMOON fluorescence spectrum can be detected in turbid non-dairy creamer solution with extinction 2.0, and through 9 mm of chicken breast tissue, suggesting that whole mouse imaging is feasible by using this strategy.

Measuring biochemical concentrations and biophysical properties within tissues is an important analytical and biomedical challenge. Fluorescence based chemical analysis techniques can be highly sensitive, versatile, and non-invasive, however, fluorescence sensing in tissue is hindered by autofluorescence, absorbance and scattering. Interference from autofluorescence can be reduced by using red and near-infrared exciting dyes or long lifetime dyes, but these characteristics significantly limit the selection of dyes. It is difficult to measure tissue autofluorescence precisely because it varies between samples and can vary in time due to changes in metabolism or bleaching. An elegant approach to *in situ* background subtraction is to use magnetically modulated probes and subtract spectra or images with probes in the "On" state from spectra or images with probes in the "Off" state. The probe spectrum can be determined by subtracting modulated spectra; the probe position can be seen by subtracting images; and the mechanical environment can be probed by measuring the rate of magnetically driven motion. Magnetic modulation of spectrochemical probes has been used for pH sensing with probes containing pH indicator dyes[1,2] and no-wash assays based on changes in probe spectrum accompanying the binding of fluorescent dyes to the probe surface in sandwich or competitive assays.[3-5] Magnetic modulation of probe position has been used in optical contrast for magnetically labeled cells,[6] laser speckle imaging contrast,[7] and optical coherence tomography contrast.[8] The rate of magnetically driven rotation has been used for viscosity sensors[2,9] and bacterial growth sensors.[10] However, the effect of scattering on magnetic modulated fluorescence has not yet been studied in detail and it is not known how deeply in tissue MagMOON fluorescence may be observed before the signal is attenuated too much by absorbance and scattering.

Tissue scattering and absorption distort fluorescence spectra due to higher transmission of red light than blue light. Light in highly scattering media can be modeled as a diffuse photon wave which decreases in intensity as it propagates.[11-13] The attenuation factor for 600-1000 nm light propagating through 1 cm of tissue is between 3 and 300, depending on the type of tissue and wavelength used.[11] If the tissue absorption and scattering properties and sample geometry are known, it is possible to use the measured spectrum to estimate tissue depth and deconvolve the original dye spectrum or lifetime.[12,13] However, this deconvolution process requires that the background autofluorescene intensity is minimal or precisely known and subtracted. MagMOONs are potentially well suited for measurement in tissue because modulation allows such *in situ* background subtraction.

MagMOONs were prepared using a previously reported method[3] with minor variations. 4.8 µm diameter Polystyrene microspheres containing both fluorescent dye (Nile red) and ferromagnetic $CrO_2$ nanorods (Spherotech, Libertyville, IL) were deposited on a microscope slide to form a monolayer. The microspheres were coated by 50 nm of silver using thermal vapor deposition at a pressure of <5x10$^{-6}$ torr (Auto 306, BOC Edward, West Sussex UK). During vapor deposition, the silver atoms travelled in straight lines from the heated source to the microspheres and thus deposited only on the side facing the source leaving shadowed side uncoated. The hemispherical silver capping layer blocked the entrance of the excitation light as well as the emission of the fluorescence from one side of the MagMOON. After coating, the microspheres were magnetized in order that all the coated/uncoated sides of the microspheres had the same magnetic dipole orientation. The MagMOONs were removed from the glass slide and suspended in deionized water by sonication. The MagMOON suspension was then loaded into a rectangular capillary (Fredrick Dimmock, Millville NJ) by capillary action and viewed with epi-fluorescence microscopy.

MagMOONs serve as compasses and align with an external magnetic field. By rotating the external field, the MagMOONs rotate in response, blinking "On" and "Off" as they rotate (see Figure 1a). Figure 1(b) shows the experimental setup to magnetically modulate MagMOONs and detect their blinking fluorescence. A hollow cylindrical NdFeB permanent magnet (½" outer diameter, ¼" inner diameter, 3" long) magnetized through its diameter (Supermagnetman, Birmingham AL) was attached to a computer-controlled stepper motor (Si-2035, Applied Motion Products, Watsonville CA). This magnet was rotated clockwise and anticlockwise 180° to turn the MagMOONs in solution beneath the magnet to "On" or "Off" orientations. The samples were observed with a DMI5000 epifluorescent microscope (Leica Microsystems, Bannockburn IL) equipped with a xenon light source, a DFC360 imaging CCD camera (Leica), and a DeltaNu DNS300 spectrometer equipped with an DUS-420A-BV CCD camera for spectroscopy (Andor Technology, South Windsor CT). MagMOONs were observed using a 10x 0.3 NA objective lens.

To measure the effect of scattering on the modulated MagMOON intensity, MagMOONs were imaged as an increasing concentration of scattering non-dairy creamer solution was added. The MagMOONs were held in a sealed capillary which was placed in a Petri dish filled with 20 mL of water. The capillary was placed on 2.3 mm aluminum pieces to fix its height above the bottom of the Petri dish. Non-dairy creamer was used as a standard scattering agent which was previously used for tissue phantoms.[14] A stock solution of non-dairy creamer was prepared by adding 0.3 g of creamer powder into 20 mL of DI water, followed by continual stirring for 1 hour. Aliquots of the stock solution were then added into the Petri dish and well mixed. After each addition, the extinction of the solution was determined by measuring light transmission in bright field microscopy mode, and the blinking signal of the MagMOONs was measured in epi-fluorescence mode using a green excitation cube (500-550 nm with 572 nm long pass emission filter, Chroma, Bellows Falls VT).

The modulated fluorescence intensity at the beginning and the end of the tests (extinction = 0 and 2.0, respectively) are shown in figure 2(a). The external magnetic field was rotated to orient the particles in a series of five On/Off pulses ("On" for 1 s, and "Off" for 1s), followed by waiting in the "Off" state for 7 s. As the concentration of scattering solution increased, less light from an overhead lightsource was transmitted and the effective extinction increased from 0 to 2 absrobance units.  After measuring extinction for transmitted light, we measured the fluorescence intensity for "On" and "Off" orientations. The modulated peak height ("On" minus "Off") decreased due to attenuation of excitation and emission light as well as reduced contrast between "On" and "Off" states. Figure 2(b) shows how the modulated signal intensity decreased as a function of the measured extinction coefficient. The value of "On" minus "Off" decreased monotonically with increasing scattering, although it decreased more rapidly for low concentrations than for high concentrations.  At the same time, the autofluorescence of the creamer solution, the average signal (defined as (On + Off)/2) increased after the addition of creamer as also shown in figure 2(b).  A continuous bleaching of the background was observed in all data.

To demonstrate imaging through tissue, MagMOONs were imaged through chicken breast tissue. Chicken breast (90% fat free, Tyson Foods Inc., Springdale AR) was cut into 3 mm sections by placing the tissue between a 3 mm spacer and sliding a sharp razor across it. The water-suspended MagMOONs were sealed into a capillary and then placed on the top of the chicken layers. The capillary was further covered by a thick layer of chicken (> 1 cm) to imitate a sample embedded within a thick tissue. The modulated MagMOON signal increased upon adding the top layer above the capillary because scattering redirects excitation light which would otherwise escape through the top surface (effectively increasing the excitation pathlength). A photo of the experimental setup is shown in figure 3(a, b). Fluorescence was measured through 3, 6, or 9 mm slices of chicken breast using a Delta-Nu DNS 300 spectrometer. A time series of 500 spectra were recorded while the MagMOONs were turned "On" and "Off" by the external magnetic field in a regular square-wave pulse train. By subtracting the average "Off" spectrum from the average "On" spectrum, the modulated MagMOON spectrum could be obtained subtracting off any unmodulated background. We calculated the magnetically modulated spectrum as the average of the first 16 "On" spectra minus the average of the 15 "Off" spectra between the "On" spectra. This algorithm accounts for and removes the essentially linear bleaching spectrum from the modulated MagMOON spectrum. Figure 4(c) shows the average spectra of "On," "Off," and "On" minus "Off" through 3 mm of chicken breast. Figure 4(d) shows the modulated MagMOON spectra after passing through 3 mm, 6 mm, and 9 mm of tissue. The modulated intensity decreased with tissue thickness, however the spectra were readily separated from tissue autofluorescence which was over 2,000 times greater in intensity for the 9 mm case and changed in time due to bleaching. To compare the spectral shapes, the MagMOON spectra were normalized and Savinsky-Golay smoothed (30 nm window) in figure 4(e). Small spectral red-shifts were observed with increasing tissue thickness, which could be attributed to reduced absorbance and scattering for red light. Comparing the normalized modulated spectrum for 3 mm and 9 mm tissue depths, after passing through 9 mm of tissue, 60% more light is attenuated at 580 nm than at 680 nm. If ratiometric indiactor MagMOONs[1,2,15] were used in place of the Nile Red MagMOONs, these spectral shifts would cause systematic errors in measured analyte concentration  A careful spectral analysis would be needed in order to determine particle depth and reconstruct the original spectrum. Temporal changes in analyte concentration could be detected without decovolution.

In conclusion, the fluorescence intensity and spectra of MagMOONs were detected through turbid media with strong scattering and autofluorescence. Nile Red dyed MagMOONs with green excitation and red-NIR emission, were detectable in concentrated creamer solution with extinction 2.0 or through at least 9 mm of chicken breast (a thickness which is approximately the "radius" of a mouse). In future work we will use pH indicator MagMOONs to measure tissue acidosis, as well as measuring biophysical properties of MagMOONs phagocytosed in alveolar macrophages (an optical version of magnetic twisting cytometry).[16] We will study the effect of absorbance in tissues with more hemoglobin, potentially requiring red-excited dyes. Ultimately, the technique is expected to be useful in non-invasive spectrochemical and biophysical sensing in small animals such as mice.

**Acknowledgement:** Funding was provided by SC-EPSCoR grant 2007068, a scholarship from VEF to KV.N, and the Clemson University startup package to J.N.A.

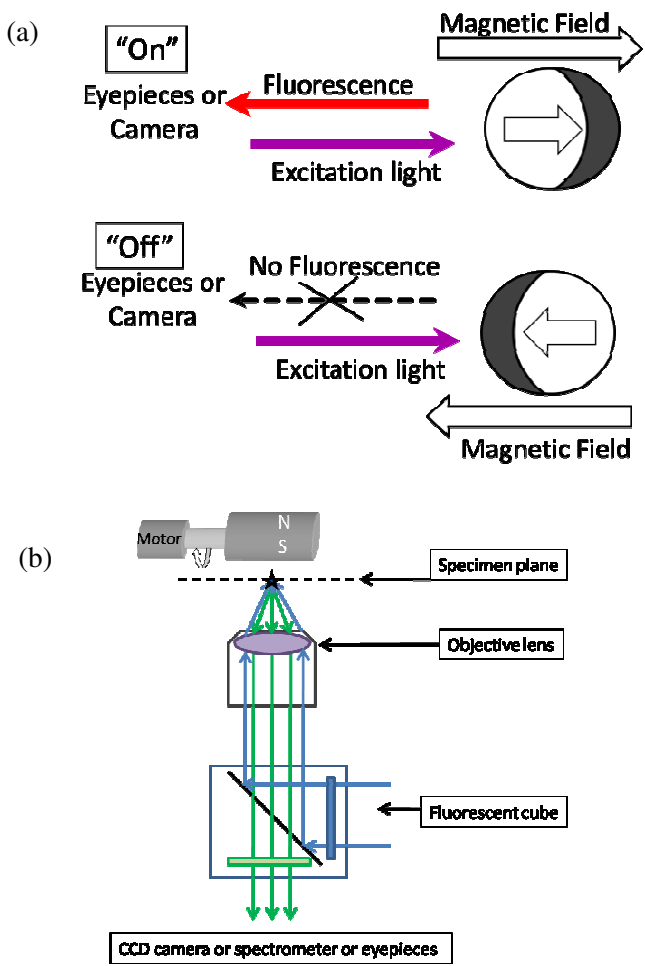

**Figure 1.** (a) Schematic showing a MagMOON magnetically controlled in "On" and "Off" orientations. (b) Experimental setup for the magnetic modulation and optical detect of MagMOONs.

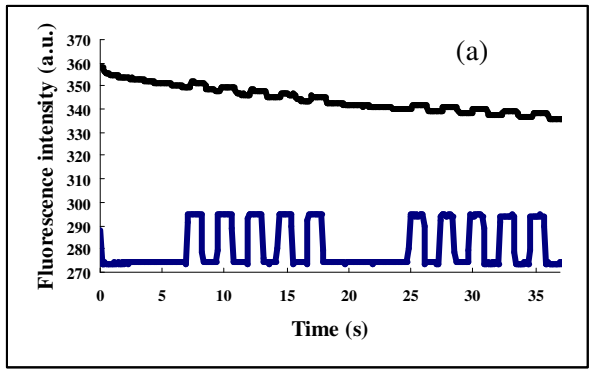

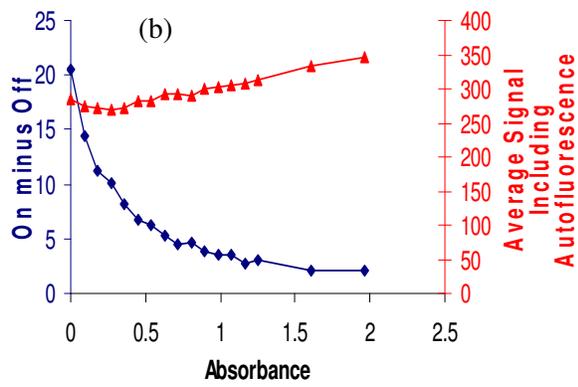

**Figure 2.** (a) Modulated fluorescence intensity when changing the extinction of the creamer solution. Blue line: extinction = 0; black line: extinction = 2.0. (b) Effect of extinction on modulated fluorescence intensity (blue curve on left axis) as well as on average intensity (red curve on right axis).

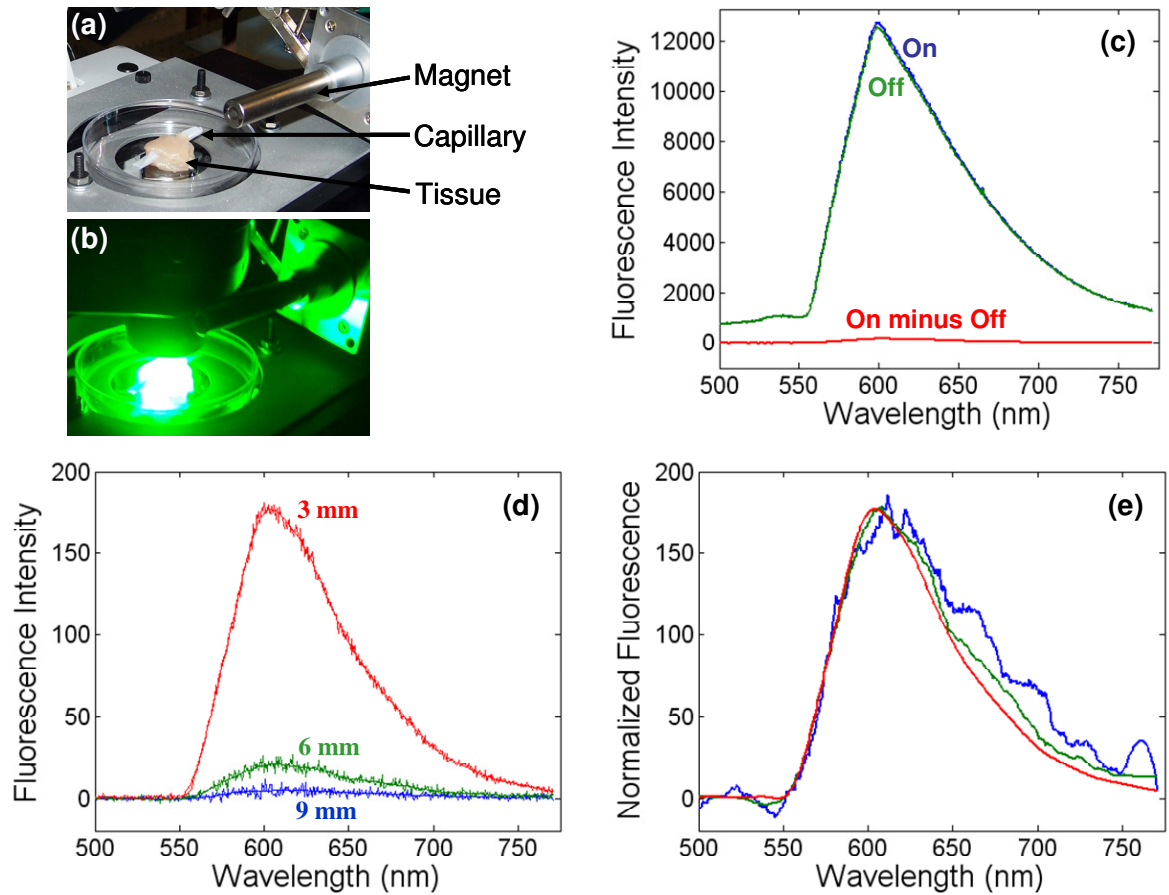

**Figure 3.** Magnetically modulated fluorescence spectra through chicken breast. (a) Photo of the experimental setup for magnetic modulation and fluorescence viewed under room light. (b) Photo viewed with the green excitation illumination. (c) Fluorescence spectra of "On," "Off," and "On minus Off" through 3 mm of chicken breast with green excitation. (d) Rescaled "On" minus "Off" spectra through 3 mm, 6 mm, and 9 mm of chicken breast. (e) Normalized and Savinsky-Golay smoothed spectra shown in (d).